# Assessment of First-principles and Semi-empirical Methodologies for Absorption and Emission Energies of $Ce^{3+}$-doped Luminescent Materials

*Yongchao Jia,*[a,b] Samuel Poncé,[c] Anna Miglio,[a,b] Masayoshi Mikami,[d] and Xavier Gonze[a,b]*

[a] Institute of Condensed Matter and Nanosciences, Université catholique de Louvain
Chemin des étoiles 8, bte L07.03.01, B-1348 Louvain-La-Neuve, Belgium;
[b] European Theoretical Spectroscopy Facility
[c] Department of Materials, University of Oxford, Parks Road, Oxford, OX1 3PH, UK
[d] MCHC R&D Synergy Centre, Inc.1000, Kamoshida-cho Aoba-ku, Yokohama, 227 8502, Japan

**Abstract:** In search of a reliable methodology for the prediction of light absorption and emission of $Ce^{3+}$-doped luminescent materials, thirteen representative materials are studied with first-principles and semi-empirical approaches. In the first-principles approach, that combines constrained density-functional theory and ΔSCF, the atomic positions are obtained for both ground and excited states of the $Ce^{3+}$ ion. The structural information is fed into Dorenbos' semi-empirical model. Absorption and emission energies are calculated with both methods and compared with experiment. The first-principles approach matches experiment within 0.3 eV, with two exceptions at 0.5 eV. In contrast, the semi-empirical approach does not perform as well (usually more than 0.5 eV error). The general applicability of the present first-principles scheme, with an encouraging predictive power, opens a novel avenue for crystal site engineering and high-throughput search for new phosphors and scintillators.

The 4f→5d transition of $Ce^{3+}$ ion has been widely used in the design of efficient luminescent systems such as white LED phosphors, scintillators and laser materials due to its spin-allowed character and its tunability as a function of the host material.[1-6] Until now, most of the efforts to find new hosts relied on trial and error. An accurate and efficient methodology to design new materials would be a remarkable achievement. With this idea in mind, a semi-empirical model has been proposed by Dorenbos, to describe $Ce^{3+}$ luminescence in inorganic compounds.[7] This semi-empirical model provides correct general trend for absorption. Unfortunately, it suffers from several drawbacks. First, its quantitative predictions rely on some fitting parameters, which have been determined only for oxide, nitride and fluoride materials at present.[6] Second, the semi-empirical model fails to predict the emission energy and Stokes shift. This limitation is due to missing experimental relaxed structure configurations in the excited state. These two shortcomings result in limited accuracy and scope of the semi-empirical approach.

In a recent paper, we have explored a first-principles alternative to overcome the drawbacks of Dorenbos' semi-empirical model.[8] The successful quantitative description of the neutral excitation, emission energy and Stokes shift in two $Ce^{3+}$-doped lanthanum silicate nitrides has been realized. Our approach is based on constrained density functional theory (CDFT) and the ΔSCF methods, following the early work by Marsman.[9] Recently, Canning and co-authors used CDFT to identify several promising hosts for efficient scintillators, but they did not study the emission and Stokes shifts.[10] In the present work, we assess the generality and accuracy of the proposed theoretical method, and compare it with Dorenbos' semi-empirical model. To do so, firstly we study from first principles the absorption, emission and Stokes shift of a set of thirteen representative $Ce^{3+}$-doped materials that include oxides, nitrides and halides, and that span a large range of transition energies, from 2 to 5 eV. Then, the first-principles structural characterization of the ground state as well as the excited state are fed into the Dorenbos' semi-empirical model. Finally, experimental transition energies and Stokes shift are compared to the ab-initio simulation and semi-empirical model to assess their accuracy and generality.

Detailed information about the Dorenbos' semi-empirical model and theoretical method can be found in the Supporting Information. The calculations in this work were performed within density functional theory (DFT) as implemented in the ABINIT package.[11-13] Still, DFT is a ground-state theory and its generalization to the excited state description thanks to CDFT does not benefit from a strong theoretical basis. The present work is thus justified by the comparison with experimental data. **Figure 1** shows the electron occupancy in CDFT. The excited state of the $Ce^{3+}$ ion is obtained by constraining the very localized predominantly 4f bands to be unoccupied, while occupying the lowest state lying higher in energy (for most cases, this state is identified afterwards to be a localized 5d state of the $Ce^{3+}$ ion). Total energy differences ($\Delta$SCF) between the ground state and such excited state will be associated to the energy change due to photon absorption and emission. The absorption and emission energy are determined at the relaxed geometry of the electronic ground and excited state, respectively, which corresponds to the Figure S1 of Supporting Information.

**Table 1** shows the transition energy and Stokes shift from our first-principles calculations and from experimental data for the thirteen representative $Ce^{3+}$ ion doped materials. The details on the geometry optimization and band structure results for all the compounds are reported in the Supporting information. **Figure 2(a)-(c)** compare first-principles and experimental results. In general, the experimental and computed values for absorption and emission energies are within 0.3 eV of each other, although in two cases, the agreement is at the level of 0.5 eV. First-principles Stokes shifts are within 30% of the experimental data, with one exception at 50%. The results for $LiYSiO_4$:Ce and $LaF_3$:Ce are less satisfactory than in other materials and might perhaps be associated with the very light Li atom (vibrational effect) and strongly electronegative F atom. Apart from these two materials, we deduce that: (1) the first-principles methodology can successfully describe oxide-, nitride and halide-based phosphors, with absorption and emission energies in the region of 2-5 eV; (2) the atomic geometry of ground and excited states for the $Ce^{3+}$-doped materials is reasonably well described. This highlights the potential of such first-principles approach for the high-throughput design of novel $Ce^{3+}$-based optical materials. Also, such first-principles approach can provide a theoretical insight into the crystal-site engineering approach, which has been recently proposed from experiment to tune the luminescence of rare earth doped phosphors.**[23]** Indeed, the luminescent center in $Ce^{3+}$ doped phosphor has been directly identified from our ab-initio method as the crystal site leading to the lowest $Ce_{5d}$ state in the bandgap of the host material.

Based on the obtained structural geometry, we extend the scope of the Dorenbos' semi-empirical model from absorption analysis to the prediction of emission energy and Stokes shift. Relying on the semi-empirical redshift D(*A*) computed for the ground and excited structural geometry, the absorption and emission energy can be defined as

$$E(A) = 49340 \text{ cm}^{-1} - D(A) \quad (1)$$

In this equation, 49340 $cm^{-1}$ is the energy of the first 4f→5d transition of $Ce^{3+}$ as a free (gaseous) ion. Stokes shift values can then be calculated as the difference between absorption and emission energies. Here, eleven compounds (oxides, nitrides, and one fluoride), among the thirteen materials used for the first-principle study, for which the spectroscopic polarization $\alpha_{sp}$ is available, have been selected for this analysis. The information needed for the determination of the redshift, D(*A*), is listed in the Supporting information. We consider the lowest and highest limits for the contribution from the crystal field splitting. The average of the high and low-limit results is compared with experiment in **Figure 2(d)-(f)** while the error bar stands for the low and high limits in the extended semi-empirical approach. The fitting result for LuAG:Ce was not included in these figures because its Stokes shift is negative.

**Figure 2(g)-(i)** shows the comparison between first-principles calculation and semi-empirical model. A statistical analysis was performed to find the linear relationship between experiment

and predictive models. The detailed analysis is given in **Table 2**. From these results, it can be concluded that, in the present domain of applicability of the Dorenbos' semi-empirical model, the first-principles approach gives generally more accurate optical transition energies and Stokes shift. Actually, the first-principles transition energies and Stokes shift can be further corrected based on the statistical analysis parameters, namely slope and intercept of the linear fit between first-principle calculation and experiment. The results are shown in the Supporting information, which give all the corrected transition energy matching the experiment within 0.3 eV. On the other hand, the predictive power of the semi-empirical method is limited by the following factors. First, the semi-empirical method provides a negative Stokes shift for LuAG:Ce, in contradiction to experiment. Second, the semi-empirical method cannot provide a correct trend for the $Ce^{3+}$ emission in $LaSi_3N_5$ and YAG. Both issues are solved in our ab-initio approach.

In summary, we analyzed the luminescence in thirteen different $Ce^{3+}$-doped materials using first-principles calculations and a semi-empirical model. The obtained results show that the first-principles approach, based on CDFT and the ΔSCF method can accurately describe the neutral excitations in these materials, and gives transition energies and Stokes shift that are generally within 0.3 eV and 30% (with two exceptions at 0.5 eV and 50%) of the experimental data, respectively. The general applicability of this method has been validated and can be used in high-throughput computational screening and crystal-site engineering of novel luminescent systems. The quantitative analysis based on Dorenbos' semi-empirical model is limited in its generality, and is less accurate than the first-principles approach. Such limitation might be ascribed to the fact that only structural information is explicitly taken into account in the semi-empirical model while the details of the electrostatic, exchange and correlation energies, linked to the (de)localization of the occupied 5d electronic level in the excited state should have an important effect on the luminescence, which is reasonably described in our first-principles calculations. We think that the latter argument explains the intriguing agreement of the CDFT and ΔSCF approach with experiment.

**Supporting Information**

Supporting Information is available from the Wiley Online Library or from the author.

**Acknowledgements**

We acknowledge the help of J.-M. Beuken for computational issues. This work, done in the framework of ETSF (project number 551 and 569), has been supported by FRS-FNRS (Belgium) and the PdR Grant No. T.0238.13 - AIXPHO. Computational resources have been provided by the supercomputing facilities of CISM/UCL and CECI funded by the FRS-FNRS under Grant No. 2.5020.11.

Received: ((will be filled in by the editorial staff))

Revised: ((will be filled in by the editorial staff))

Published online: ((will be filled in by the editorial staff))

**Figures and Tables**

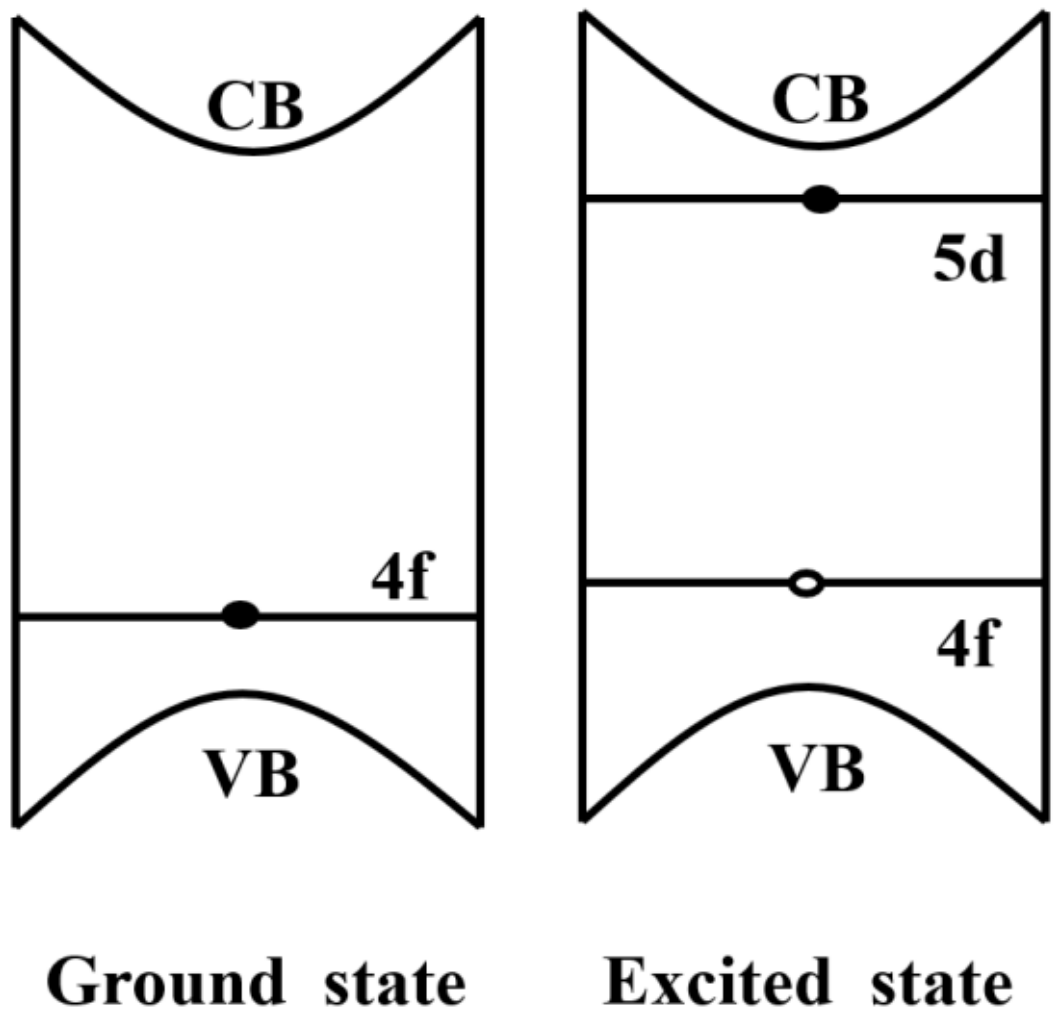

**Figure 1.** Electron occupation of the levels inside the host band gap, in the ground state and excited state.

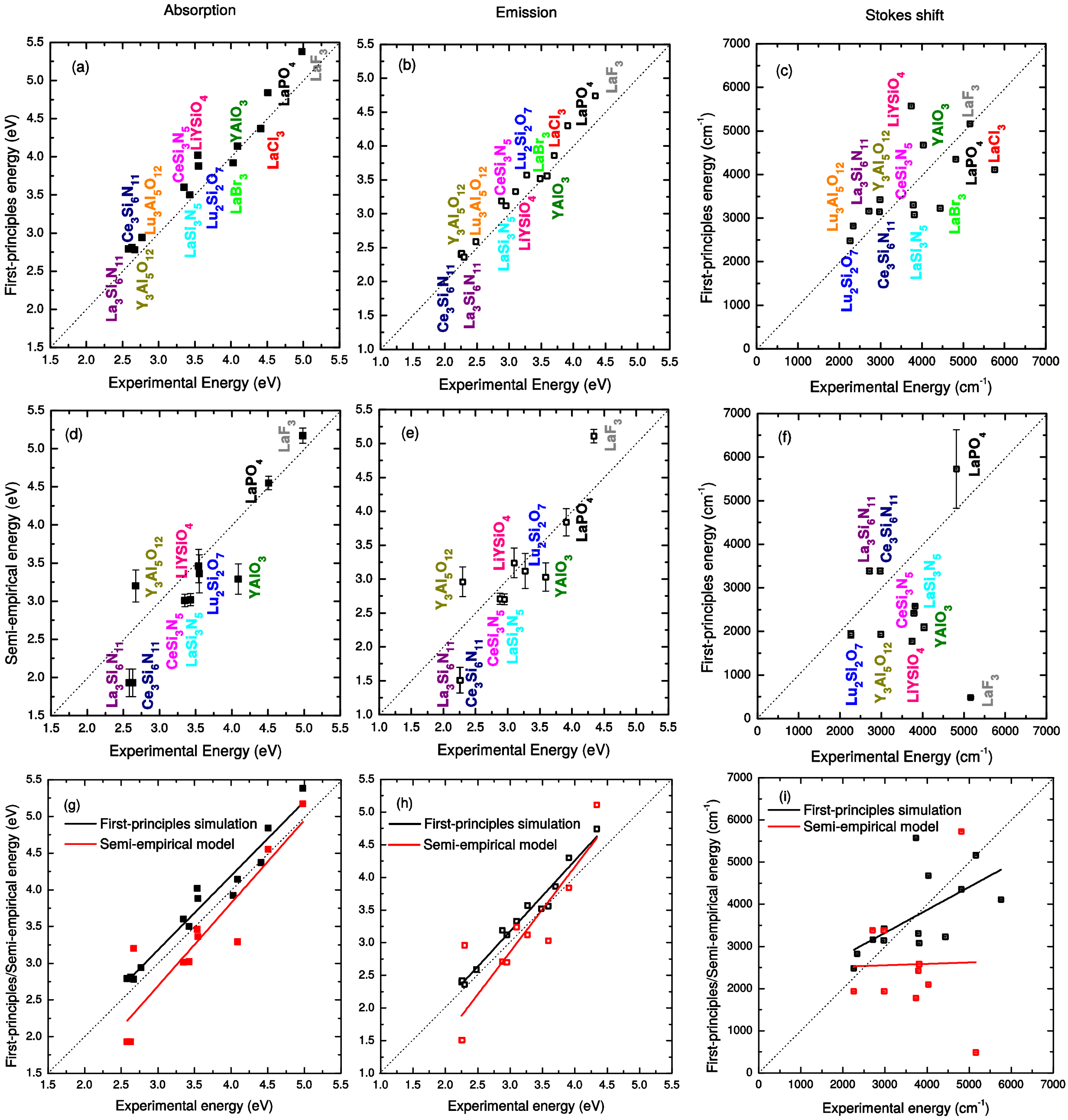

**Figure 2.** Comparison between experimental results and first-principles calculations: (a) absorption (b) emission and (c) ΔS; between experimental results and semi-empirical model for (d) absorption, (e) emission, and (f) ΔS, (g) fitting line of absorption energy, (h) fitting line of emission energy, (i) fitting line of ΔS. The error bars for the semi-empirical model stem from the low and high limits of the crystal-field splitting.

**Table 1.** Absorption (Abs, eV), emission (Em, eV) energy and Stokes shift ΔS ($cm^{-1}$), from first-principles calculations and experiment for thirteen $Ce^{3+}$ doped host materials. Only few numbers in bold deviate substantially from experiment.

| | First-principles | | | Experiment | | | |
|---|---|---|---|---|---|---|---|
| Compounds | Abs | Em | ΔS | Abs | Em | ΔS | Ref |
| $La_3Si_6N_{11}$:Ce | 2.79 | 2.40 | 3160 | 2.58 | 2.25 | 2717 | [14] |
| $Ce_3Si_6N_{11}$ | 2.81 | 2.42 | 3146 | 2.63 | 2.26 | 2974 | [15] |
| $Y_3Al_5O_{12}$:Ce | 2.78 | 2.36 | 3424 | 2.67 | 2.30 | 2984 | [16] |
| $Lu_3Al_5O_{12}$:Ce | 2.94 | 2.59 | 2823 | 2.77 | 2.48 | 2339 | [16] |
| $CeSi_3N_5$ | 3.60 | 3.19 | 3307 | 3.35 | 2.88 | 3791 | [17] |
| $LaSi_3N_5$:Ce | 3.50 | 3.12 | 3080 | 3.43 | 2.95 | 3815 | [18] |
| $LiYSiO_4$:Ce | **4.02** | 3.33 | **5575** | 3.54 | 3.10 | 3740 | [19] |
| $Lu_2Si_2O_7$:Ce | 3.88 | 3.57 | 2480 | 3.55 | 3.27 | 2258 | [20] |
| $LaBr_3$:Ce | 3.92 | 3.52 | 3226 | 4.03 | 3.48 | 4439 | [21] |
| $YAlO_3$:Ce | 4.14 | 3.56 | 4678 | 4.09 | 3.59 | 4033 | [7a] |
| $LaCl_3$:Ce | 4.37 | 3.86 | 4113 | 4.41 | 3.70 | 5762 | [22] |
| $LaPO_4$:Ce | 4.84 | 4.30 | 4355 | 4.51 | 3.91 | 4818 | [7a] |
| $LaF_3$:Ce | **5.38** | **4.74** | 5162 | 4.98 | 4.34 | 5162 | [7b] |

**Table 2.** Statistical analysis of transition energy and Stokes shift from first-principles calculation and semi-empirical model. ME (eV, cm-1), MAE (eV, cm-1), MRE (%) and MARE (%) stand for the mean error, mean absolute error, mean relative error, and mean absolute relative error, respectively. The slope, intercept and coefficient of determination ($R^2$) correspond to the linear fitting in Figure 2(g)-(i). The most problematic quantities are indicated in bold.

| | First-principles | | | Semi-empirical | | |
|---|---|---|---|---|---|---|
| | Abs | Em | ΔS | Abs | Em | ΔS |
| ME | 0.175 | 0.205 | 33.5 | -0.118 | 0.027 | **-1118** |
| MAE | 0.205 | 0.210 | **728** | 0.350 | 0.423 | **1502** |
| MRE | 5.100 | 6.280 | 4.17 | 3.540 | -0.01 | **-26.7** |
| MARE | 5.850 | 6.410 | **19.1** | **10.8** | **14.6** | **37.3** |
| Slope | 1.010 | 1.080 | 0.543 | 1.13 | 1.30 | **0.033** |
| Intercept | 0.142 | -0.066 | 1677 | -0.713 | -1.041 | 2453 |
| $R^2$ | 95.1 | 97.1 | 33.3 | 81.4 | 76.6 | **-12.4** |

**Entry for the Table of Contents**

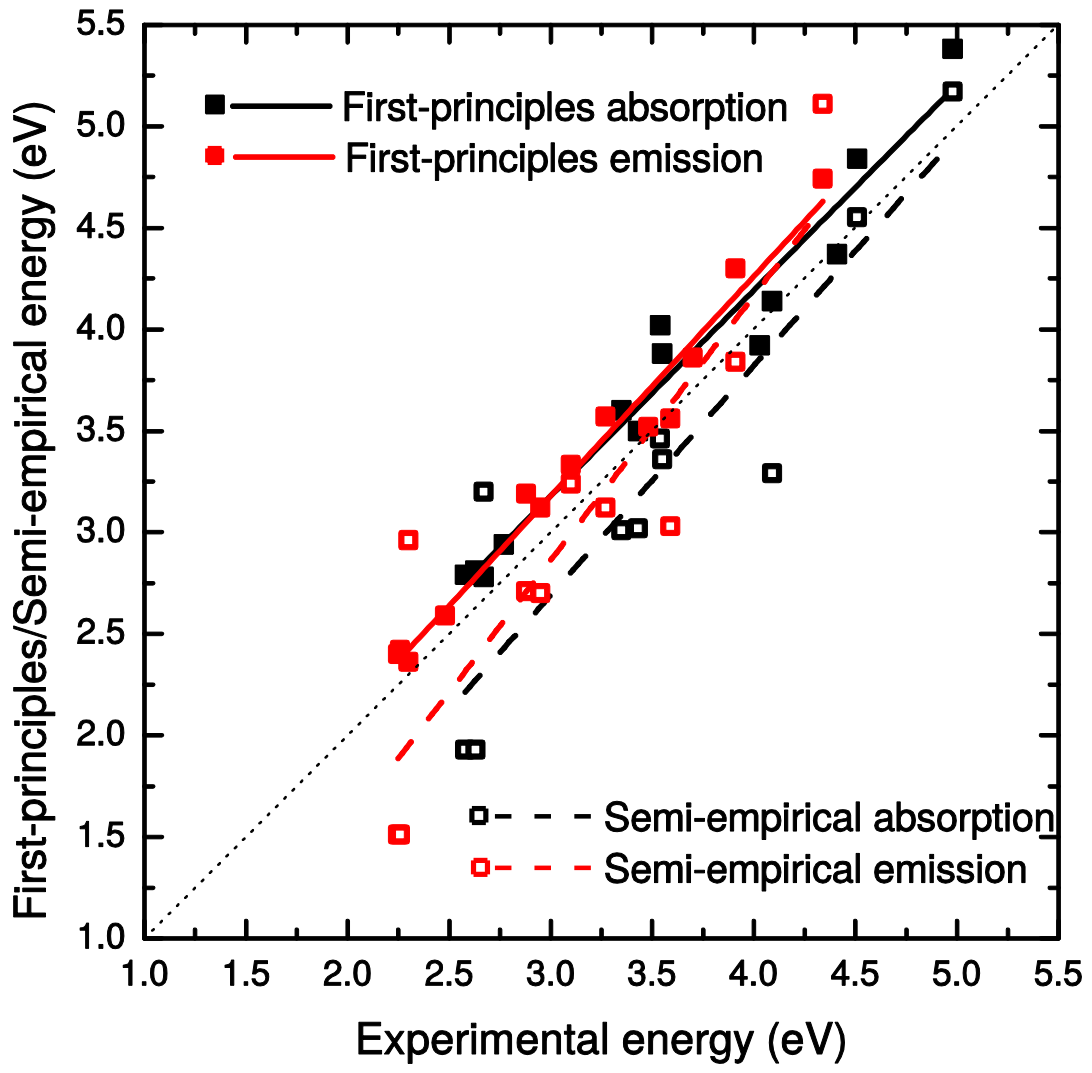

**Light absorption and emission of thirteen $Ce^{3+}$-doped materials** were studied from first-principles and Dorenbos' semi-empirical model. For both methods, the obtained transition energies and Stokes shift are compared to the values from experiment. The statistical analysis shows that the first-principles calculation gives a better consistency than the semi-empirical model.